# Solving compressible Navier-Stokes equations using the feature-enhanced neural network

Jiahao Song[1,2,3], Wenbo Cao[1,2,3], Weiwei Zhang[1,2,3,*]

[1] School of Aeronautics, Northwestern Polytechnical University, Xi'an 710072, China

[2] International Joint Institute of Artificial Intelligence on Fluid Mechanics, Northwestern Polytechnical University, Xi'an, 710072, China

[3] National Key Laboratory of Aircraft Configuration Design, Xi'an 710072, China

* Corresponding author. E-mail: aeroelastic@nwpu.edu.cn

**Abstract** Physics-informed neural networks (PINNs) have shown remarkable prospects in solving partial differential equations (PDEs) involving fluid mechanics. However, the method has so far succeeded only in inviscid flows and incompressible viscous flows, while the solution of compressible viscous flows still faces significant challenges. In previous work, we proposed a feature-enhanced neural network (FENN), which enhances the ability of PINNs to approximate flows by introducing beneficial features into the network inputs, thereby improving the performance in solving PDEs. In this study, we extend FENN to compressible viscous flows, which are governed by the compressible Navier-Stokes equations including the continuity, momentum, and energy equations. By solving four forward problems under different flow conditions and geometries together with a parametric problem involving angle of attack, we validate the effectiveness of FENN. In contrast, existing advanced methods that are well established for inviscid flows and incompressible viscous flows fail in this scenario. To the best of our knowledge, this is the first time that a PINN-like method has successfully solved forward and parametric problems involving compressible viscous flows.

## 1. Introduction

Driven by recent advances in computer technology and machine learning, deep neural networks have become a popular tool for solving partial differential equations (PDEs). Representative methods include physics-informed neural networks (PINNs) [1], the deep Ritz method (DRM) [2], and the deep Galerkin method (DGM) [3], among which PINNs have attracted the greatest research interest. PINNs integrate the PDE residuals into the loss function to ensure that the network outputs satisfy the physical constraints while approximating the boundary conditions. This idea originated in the 1990s, when several methods for solving PDEs based on neural networks were proposed [4-6]. Compared with traditional numerical methods, PINNs offer several advantages, including eliminating the need for mesh generation and enabling the solution of inverse problems [7, 8] as well as parametric problems [9, 10]. PINNs have yielded notable results in diverse fields, such as fluid mechanics [7, 11, 12], electromagnetic propagation[13, 14], heat transfer[15, 16], and materials [17, 18].

Focusing on fluid mechanics, Raissi et al. [7] employed PINNs to reconstruct the pressure and velocity fields using only concentration field, demonstrating the capability of this approach in solving inverse problems. Mao et al. [19] solved inviscid supersonic problems using PINNs and achieved better performance by refining collocation points near shock waves. Liu et al. [20] constructed a weight based on velocity gradients for the PDE residuals in the loss function to mitigate residuals near shock waves, thereby improving the performance of PINNs on shock capturing problems. Wassing et al. [21] attempted to solve inviscid transonic flows around an airfoil involving shock waves through the incorporation of artificial viscosity into the Euler equations, but their results exhibited significant shock capturing errors. Jin et al. [22] developed an incompressible Navier-Stokes equation solver NSFnets, and demonstrated its effectiveness through extensive numerical experiments. Rao et al. [23] showed the superiority of using continuum-mechanics-based equations over standard incompressible Navier-Stokes equations as PDE constraints for PINNs. Song et al. [24] identified the ill-conditioning of the PDE loss function in PINNs caused by non-uniform collocation points, and proposed a volume weighting method for the PDE residuals to alleviate this issue. Building upon this, Cao et al. [25] further introduced mesh transformation to solve inviscid subsonic flows around an airfoil. Similarly, Zhang et al. [26] combined coordinate transformation to solve incompressible turbulent flows under the condition of frozen Reynolds stresses. Chuang et al. [27] solved the unsteady incompressible viscous flow around a cylinder and observed an interesting phenomenon: PINNs obtained an unstable steady solution rather than the unsteady flow characterized by unstable vortex shedding. Huang et al. [28] employed PINNs to solve incompressible high Reynolds number boundary layer flows, and improved the accuracy using a domain decomposition approach. Qiu et al. [29] addressed two-phase flows involving reversed single vortex and bubble rising by coupling the incompressible Navier-Stokes equations with the Cahn-Hilliard equation as PDE constraints for PINNs. The solution of the lid-driven cavity flow has also attracted extensive attention [30-32].

In addition, researchers have recently studied parametric problems in fluid mechanics using PINNs. Specifically, PINNs allow the introduction of state parameters (e.g., Reynolds number and Mach number) and geometric parameters (e.g., class-shape transform parameters) into the inputs, thereby extending the dimensionality of the independent variables for PDEs. This capability enables PINNs to obtain all flows within a given space of state and geometric parameters through a single training. In contrast, traditional numerical methods are limited to solving the flow for one state and geometry in each run. Consequently, they lead to high computational cost in engineering problems, such as aircraft design, where simulating the flow under numerous states and geometries is required. Sun et al. [9] were the first to introduce fluid viscosity and geometric parameters into the inputs of PINNs to solve parametric problems of incompressible viscous blood flow. Wassing et al. [10] addressed parametric problems involving Mach number and geometric parameters for inviscid oblique shock and inviscid flow around a cylinder. Cao et al. [33] constructed a 16-dimensional parametric problem for inviscid subsonic flow around an airfoil by introducing angle of attack, Mach number, and class-shape transform (CST) parameters into the inputs of PINNs, thereby obtaining flows across an extensive space of states and geometries.

These studies have advanced the development of PINNs and expanded their applications in fluid dynamics. However, they mainly focus on compressible inviscid and incompressible viscous flows, overlooking compressible viscous flows. Solving compressible viscous flows is a key step in applying PINN-like methods to engineering problems in fluid dynamics. Therefore, this study employs the feature-enhanced neural network (FENN) proposed in our previous work [34] to solve compressible viscous flows. FENN enhances the capability of PINNs to approximate flows by introducing features highly correlated with the flows into the network inputs, thereby improving solution performance. To the best of our knowledge, this is the first time that a PINN-like method has successfully solved forward and parametric problems involving compressible viscous flows. The remainder of the paper is organized as follows. Section 2 presents the problem setting and provides an overview for PINNs. Subsequently, the feature-enhanced neural network is introduced in detail. In Section 3, FENN is validated by solving the forward and parametric problems involving compressible viscous flows around an airfoil. Finally, Section 4 offers the conclusions and outlines potential directions for future research.

## 2. Methodology

*2.1 Problem setting*

Compressible viscous flows are an important subject of study in fluid mechanics, with dynamics governed jointly by compressibility and viscosity. Compared with incompressible flows and inviscid flows, compressible viscous flows are more relevant to engineering problems, and both the flows and their governing equations are of higher complexity. In this study, we consider the dimensionless steady two-dimensional compressible Navier-Stokes equations (1).

$$\frac{\partial(\rho u)}{\partial x}+\frac{\partial(\rho v)}{\partial y}=0$$

$$\frac{\partial(\rho u^2)}{\partial x}+\frac{\partial(\rho uv)}{\partial y}+\frac{\partial p}{\partial x}-\frac{1}{Re}(\frac{\partial \tau_{xx}}{\partial x}+\frac{\partial \tau_{yx}}{\partial y})=0$$

$$\frac{\partial(\rho uv)}{\partial x}+\frac{\partial(\rho v^2)}{\partial y}+\frac{\partial p}{\partial y}-\frac{1}{Re}(\frac{\partial \tau_{xy}}{\partial x}+\frac{\partial \tau_{yy}}{\partial y})=0 \quad (1)$$

$$\frac{\partial[u(E+p)]}{\partial x}+\frac{\partial[v(E+p)]}{\partial y}-\frac{1}{Re}[\frac{\partial(u\tau_{xx}+v\tau_{xy})}{\partial x}+\frac{\partial(u\tau_{yx}+v\tau_{yy})}{\partial y}+$$

$$\frac{1}{(\gamma-1)Ma^2 Pr}(\frac{\partial^2 T}{\partial x^2}+\frac{\partial^2 T}{\partial y^2})]=0$$

where $\rho$ denotes the density. $u$ and $v$ are the $x$ and $y$ components of the velocity vector $V$. $p$, $T$ and $E$ represent pressure, temperature and total energy per unit volume, respectively. $\tau_{xx}$, $\tau_{xy}$ and $\tau_{yy}$ represent the shear stresses in each direction. $Re$ is Reynolds number, $Ma$ is Mach number, $Pr=0.72$ is Prandtl number, $\gamma=1.4$ denotes the specific heat ratio. The variables satisfy the following relations,

$$E=\frac{p}{\gamma-1}+\frac{1}{2}\rho(u^2+v^2) \quad (2)$$

$$\tau_{xx}=-\frac{2}{3}\mu(\frac{\partial u}{\partial x}+\frac{\partial v}{\partial y})+2\mu\frac{\partial u}{\partial x}$$

$$\tau_{yy}=-\frac{2}{3}\mu(\frac{\partial u}{\partial x}+\frac{\partial v}{\partial y})+2\mu\frac{\partial v}{\partial y} \quad (3)$$

$$\tau_{xy}=\tau_{yx}=\mu(\frac{\partial v}{\partial x}+\frac{\partial u}{\partial y})$$

$$p=\frac{\rho T}{\gamma Ma^2} \quad (4)$$

where $\mu$ represents the dimensionless dynamic viscosity. By substituting Eqs. (2)-(4) into Eq. (1), a closed system of equations $\mathcal{R}[U] = 0$ for the solution vector $U = [\rho, u, v, T]$ can be constructed. The boundary conditions are denoted by $\mathcal{B}[U] = 0$. For the compressible viscous flows around an airfoil considered in this study, the boundary conditions include the far-field boundary condition $[\rho_\infty, u_\infty, v_\infty, T_\infty] = [1, \cos(\alpha), \sin(\alpha), 1]$ and the wall boundary condition $u = 0$, $v = 0$, $\partial T / \partial \boldsymbol{n} = 0$ (the wall is assumed to be adiabatic), where $\alpha$ denotes the angle of attack and $\boldsymbol{n}$ represents the outward normal direction of the wall.

*2.2 Physics-informed neural networks*

For the dimensionless steady two-dimensional compressible Navier-Stokes equations (1) considered in this study, PINNs take the space coordinates $(x, y)$ as the inputs and output the approximate solution $U_\theta$. The result of PINNs is determined by the network parameters $\theta$, which are optimized by minimizing the loss function (5).

$$\mathcal{L} = \lambda_{bc} \mathcal{L}_{bc} + \lambda_r \mathcal{L}_r \tag{5}$$

where

$$\mathcal{L}_{bc} = \frac{1}{N_{bc}} \sum_{i=1}^{N_{bc}} \left| \mathcal{B}[U_\theta(x_{bc}^i, y_{bc}^i)] \right|^2 \tag{6}$$

$$\mathcal{L}_r = \frac{1}{N_r} \sum_{i=1}^{N_r} \left| \mathcal{R}[U_\theta(x_r^i, y_r^i)] \right|^2 \tag{7}$$

in Eqs. (6)-(7), $\{x_{bc}^i, y_{bc}^i\}_{i=1}^{N_{bc}}$ are training points for the boundary conditions, the corresponding boundary loss $\mathcal{L}_{bc}$ ensures the outputs of PINNs approach the boundary conditions. $\{x_r^i, y_r^i\}_{i=1}^{N_r}$ are collocation points, the corresponding PDE loss $\mathcal{L}_r$ constrains the outputs to satisfy the PDEs. The partial derivatives in $\mathcal{B}[U_\theta(x_{bc}^i, y_{bc}^i)]$ and $\mathcal{R}[U_\theta(x_r^i, y_r^i)]$ are computed via automatic differentiation [35]. $\lambda_{bc}$ and $\lambda_r$ are weighting factors used to balance the two loss, thereby ensuring a reasonable optimization. In this study, Eq. (7) is replaced with the volume weighting PDE loss function (8) introduced by Song et al. [24]. Extensive numerical experiments [21, 24-26] have demonstrated that the latter effectively alleviates the ill-conditioning of PINNs caused by non-uniform collocation points. In Eq. (8), $s(x_r^i, y_r^i)$ represents the volume occupied by $(x_r^i, y_r^i)$ in the computational domain.

$$\mathcal{L}_r = \sum_{i=1}^{N_r} \left| \mathcal{R}[U_\theta(x_r^i, y_r^i)] s(x_r^i, y_r^i) \right|^2 \bigg/ \sum_{i=1}^{N_r} [s(x_r^i, y_r^i)]^2 \tag{8}$$

## 2.3 Feature-enhanced neural network

In previous work [34], we proposed a feature-enhanced neural network (FENN). This method enhances the capability of PINNs to approximate flows by designing features highly correlated with the flows and introducing them into the network inputs, thereby improving the solution performance. Figure 1 shows a schematic of FENN for solving steady two-dimensional PDEs. $F(x,y)$ represents the features, and $K$ represents their number.

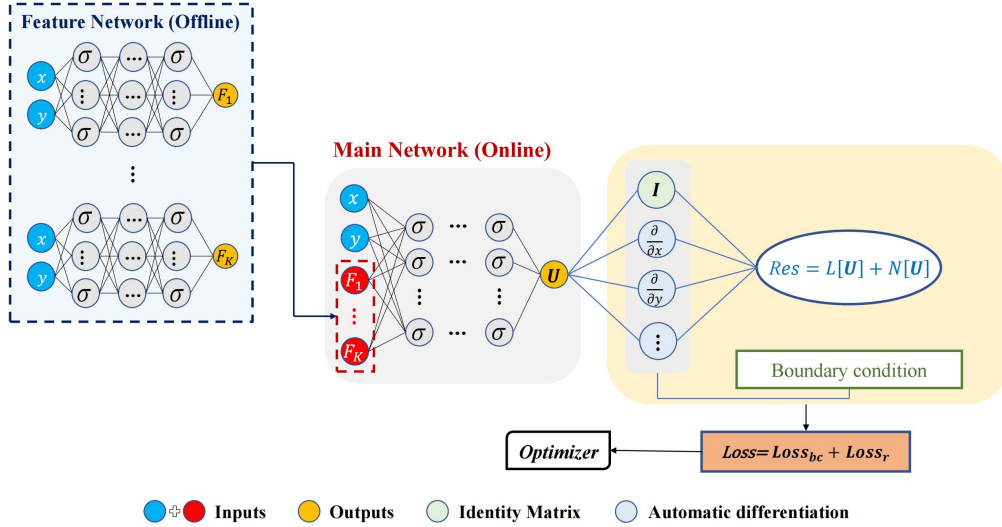

Figure 1. A schematic of FENN for solving steady two-dimensional PDEs.

The feature networks are established offline to address the incorrect computation of partial derivatives caused by directly introducing features into the inputs. We illustrate this issue by computing $\partial U_\theta / \partial x$. Since $F_j$ is a function of $(x,y)$, according to the chain rule,

$$\frac{\partial U_\theta}{\partial x} = \left(\frac{\partial U_\theta}{\partial x}\right)_s + \sum_{j=1}^{K} \frac{\partial U_\theta}{\partial F_j} \frac{\partial F_j}{\partial x} \tag{9}$$

where $(\partial U_\theta / \partial x)_s$ and $\partial U_\theta / \partial F_j$ denote the partial derivatives of the output $U_\theta$ with respect to the inputs $x$ and $F_j$, computed via automatic differentiation. Directly introducing the features into the inputs of PINNs implies that $x$ and $F_j$ are independent, which leads to $\partial F_j / \partial x = 0$. Consequently, an erroneous $\partial U_\theta / \partial x$ is obtained. To resolve this problem, the feature networks are trained offline in a supervised manner to obtain regression models of $F_j$ with respect to $(x,y)$, which are subsequently employed as features in the inputs. During the computation of $\partial U_\theta / \partial x$ in the main network, $\partial F_j / \partial x$ is extracted from the feature networks via automatic differentiation, thereby ensuring the correctness of $\partial U_\theta / \partial x$.

In [34], we designed the geometric and physical features. The former includes the shortest distance $D$ from the collocation points to the wall and the angle $\Phi$ defined based on the wall tangent, while the latter includes the pressure field and velocity field obtained by solving the potential flow equation. In this study, $D$ is selected as the feature of FENN. On the one hand, based on our experience, it plays a dominant role in geometric features compared with the angle. On the other hand, as demonstrated in [34], the enhancement effect brought by geometric features is almost the same as that of physical features, but the computational cost (including the cost of training the feature networks) of the former is lower and can be neglected in the total training time of FENN. In addition, as in PINNs, the standard PDE loss (7) is also replaced by the volume weighting PDE loss (8) in FENN.

## 3. Results

We validate the effectiveness of FENN in solving compressible Navier-Stokes equations through four forward problems under different flow conditions and geometries, together with a parametric problem involving angle of attack. For all the problems, tanh is selected as the activation function, and the network parameters are initialized using the Xavier normal initialization [36]. Unless stated otherwise, optimization is performed with the Limited-memory Broyden-Fletcher-Goldfarb-Shanno (L-BFGS) algorithm [37], where the maximum number of inner iterations per epoch is set to 20. The feature networks are constructed with 3 hidden layers, each containing 20 neurons, and are trained for 1000 epochs. Their convergence histories are provided in Appendix A.

The relative $L_2$ error (10) is computed to evaluate accuracy,

$$\frac{\|\boldsymbol{U}_\theta - \boldsymbol{U}\|_2}{\|\boldsymbol{U}\|_2} \tag{10}$$

the reference solution $\boldsymbol{U}$ is obtained using the finite volume method (FVM). PINNs and FENN are implemented using PyTorch and executed on an NVIDIA GeForce RTX 3080 GPU.

*3.1 Forward problems involving diverse flow conditions and airfoils*

We first solve compressible viscous flows around an airfoil under four diverse flow conditions and geometries. The angles of attack, Mach numbers, Reynolds numbers, and airfoils are listed in Table 1. Three classical NACA airfoils with different camber and thickness are selected to cover the typical variations in airfoil geometry, as shown in Figure 2. In Case 2, the angle of attack is set to 20° to evaluate the capability of FENN in computing separated flow.

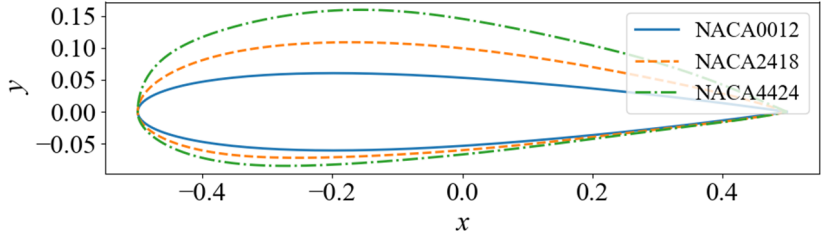

Figure 2. A schematic of three NACA airfoils.

Taking the NACA0012 airfoil as an example, Figure 3 illustrates the computational domain and collocation points (for all the cases, the far-field boundary and the number of collocation points are the same). $N_r = 20100$ collocation points are selected for computing the PDE loss. On both the far-field boundary and the airfoil, $N_{bc} = 201$ points are selected each to enforce the boundary conditions. We solve the four flows using PINNs and FENN, respectively. All the networks are constructed with 7 hidden layers, each containing 64 neurons, and are trained for 6000 epochs. Considering the properties of the volume weighting PDE loss, we set $\lambda_r = 2 \times 10^5$ and $\lambda_{bc} = 1$ following [25].

Table 1 Flow conditions and geometries for four cases.

|  | $\alpha$ | $Ma$ | $Re$ | Airfoil |
| --- | --- | --- | --- | --- |
| Case1 | 4 | 0.7 | 400 | NACA0012 |
| Case2 | 20 | 0.7 | 400 | NACA0012 |
| Case3 | -3 | 0.5 | 600 | NACA2418 |
| Case4 | -5 | 0.6 | 800 | NACA4424 |

Figure 4 shows the convergence histories of the loss functions. We observe that both PINNs and FENN fully converge. Compared with PINNs, FENN exhibits faster convergence. Moreover, both the boundary and PDE losses obtained by FENN are approximately one order of magnitude lower than those of PINNs. Figure 5 illustrates the pressure coefficients on the airfoils obtained by PINNs and FENN. Clearly, the results of PINNs exhibit large discrepancies compared with the reference solution, indicating that although PINNs using the volume weighting PDE loss (8) are capable of solving Euler equations and incompressible Navier-Stokes equations [24, 34], their applicability to compressible Navier-Stokes equations is poor. In contrast, the results of FENN show a small deviation from the reference solution. Furthermore, we plot the streamlines of the separated flow induced by a large angle of attack (Case 2). We observe that the separation vortices obtained by FENN are in good agreement with the reference solution.

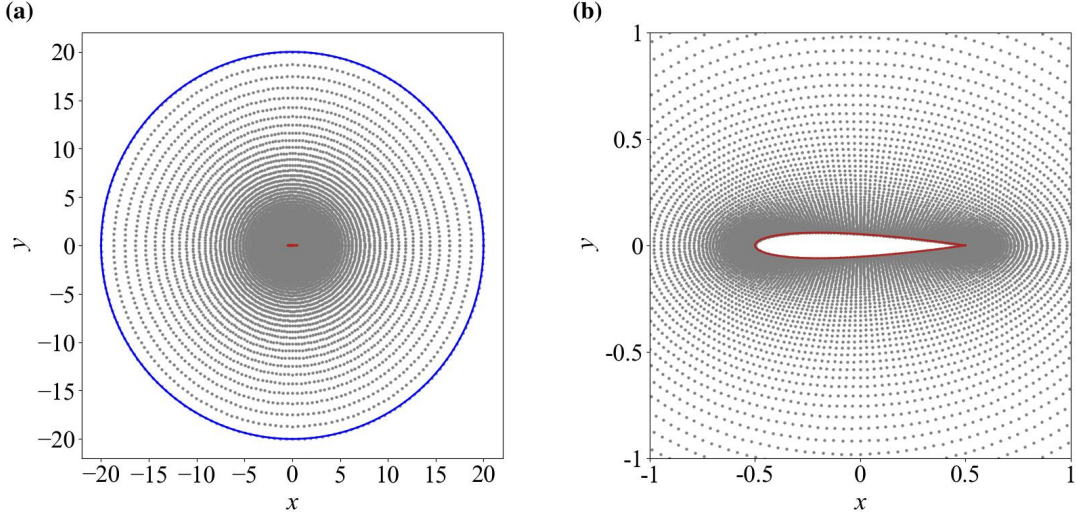

Figure 3. Computational domain and collocation points for solving compressible viscous flow around an airfoil. The blue line denotes the far-field boundary, and the red line denotes the airfoil.

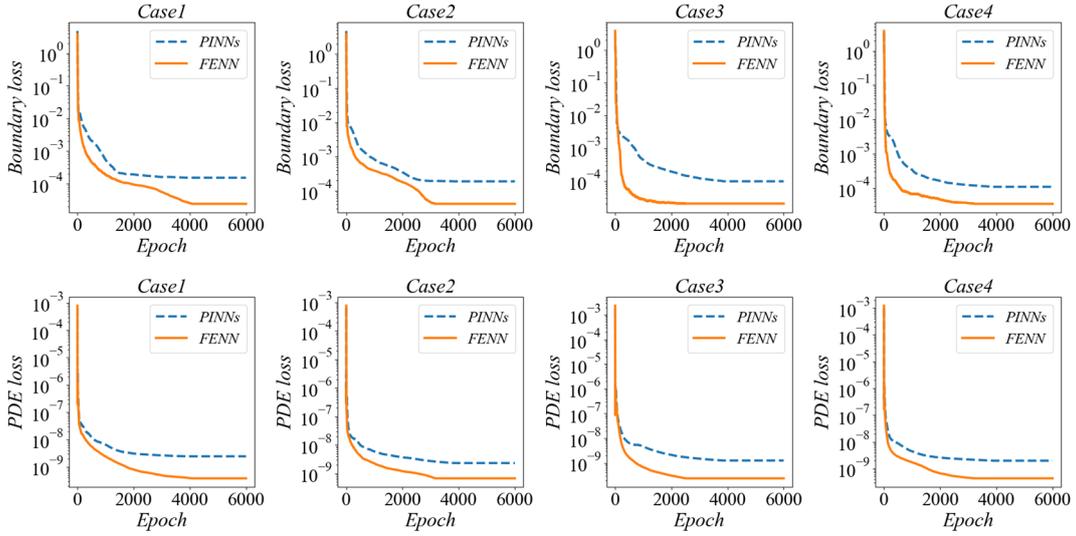

Figure 4. Convergence histories of PINNs and FENN for solving the forward problems involving compressible viscous flows around an airfoil.

*3.2 Parametric problem involving angle of attack*

In aerodynamics, it is often necessary to evaluate the pressure distribution of an airfoil at multiple angles of attack. In traditional numerical methods, the flow for each angle of attack needs to be solved individually, leading to an expensive and repetitive task. By contrast, PINNs and FENN allow the extension of the independent variables of the compressible Navier-Stokes equations from $[x, y]$ to $[x, y, \alpha]$ by introducing the angle of attack $\alpha$ into the network inputs, which is defined as a parametric problem [9, 10]. In this scenario, the angle of attack in the PDE is no longer a constant but a predefined range. Therefore, PINNs and FENN are able to generate all flows within the given angle of attack range through a single training, which is a major advantage of PINN-like methods over traditional numerical methods.

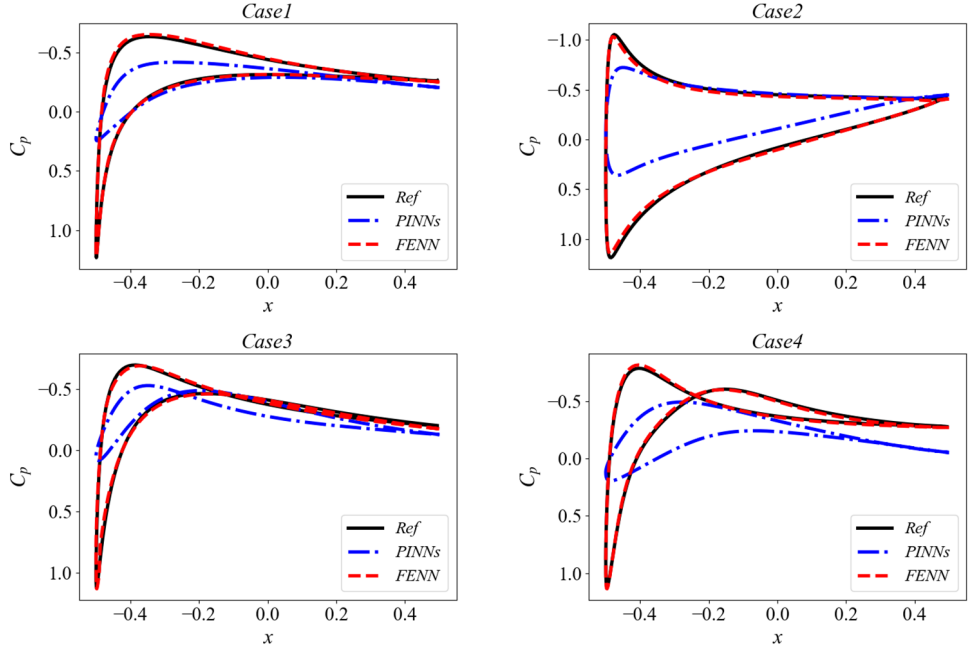

Figure 5. Pressure coefficients on the airfoils obtained by different methods for solving the forward problems involving compressible viscous flows around an airfoil.

We consider the NACA0012 airfoil, $Ma = 0.4$, $Re = 1000$ and $\alpha \in [-5°, 5°]$. The computational domain in the $(x, y)$ space and the boundary conditions for this problem are the same as those in Section 3.1. Since solving parametric problems is more challenging than solving forward problems, the performance of PINNs in the former is generally worse than in the latter [21, 33, 34]. Considering the poor performance of PINNs on forward problems (Section 3.1), we employ only FENN to solve this problem. The network is constructed with 9 hidden layers, each containing 64 neurons, and is trained for 1500 epochs. The L-BFGS algorithm is used for optimization, with a maximum of 1000 inner iterations per epoch. In each epoch, collocation points are resampled and the optimizer is reinitialized to avoid the incompatibility issue between the LBFGS optimizer and batching, thereby alleviating the requirement for extensive sampling of collocation points [33]. The number of collocation points is $N_r = 20100$, and $N_{bc} = 3000$ points are selected each to enforce the boundary conditions. We set $\lambda_r = 2 \times 10^5$ and $\lambda_{bc} = 4$ (compared with the forward problems, $\lambda_{bc}$ is increased to strengthen the boundary condition constraints primarily determined by the angle of attack).

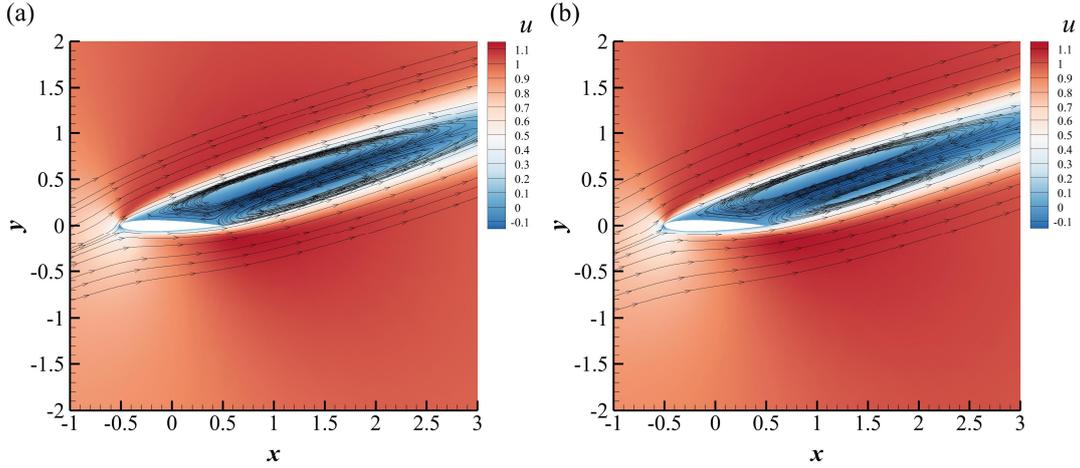

Figure 6. Streamlines near the airfoil obtained by (a) FVM and (b) FENN for solving the compressible viscous flow around the NACA0012 airfoil (Case2).

Figure 7 shows the convergence history of the loss functions. To evaluate the accuracy of FENN, we compare the pressure coefficients on the NACA0012 airfoil obtained by FENN at four different angles of attack with the reference solutions obtained by FVM (solving the flow sequentially at each angle of attack, with a total of four runs), as illustrated in Figure 8. The results of FENN exhibit small deviations from the reference solutions, demonstrating its effectiveness in solving parametric problems involving compressible viscous flows around an airfoil.

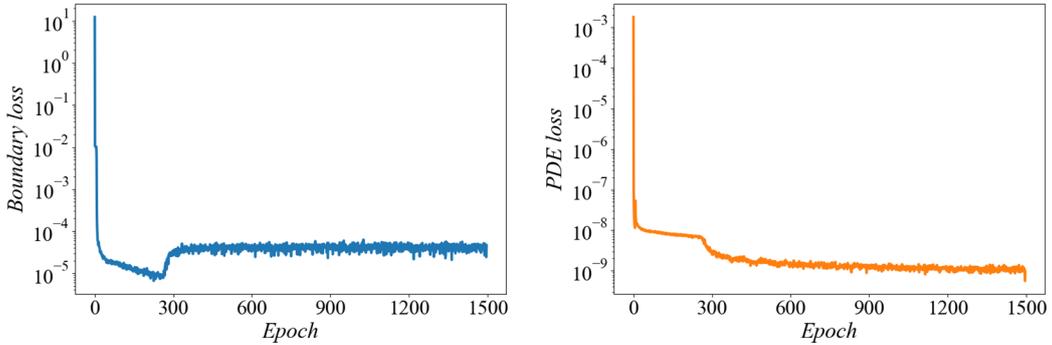

Figure 7. Convergence history of the loss functions obtained by FENN for solving the parametric problem involving compressible viscous flows around the NACA0012 airfoil.

## 4. Conclusions

In this work, we employ the feature-enhanced neural network previously proposed by us to solve compressible Navier-Stokes equations. To reduce computational cost, we prune the geometric and physical features in FENN. The shortest distance from the collocation points to the airfoil is selected as the feature, owing to its efficiency and low acquisition cost. By solving forward problems involving four different flow conditions and geometries, including one case with separation induced by a large angle of attack, we first validate the effectiveness of FENN in simulating a single flow. In contrast, PINNs using the volume weighting PDE loss fail in this scenario, although they have been successful in compressible inviscid and incompressible viscous flows. Furthermore, we solve a parametric problem involving angle of attack to demonstrate the unique advantages of FENN over traditional numerical methods. The results are likewise satisfactory. Our work demonstrates the capability of PINN-like methods for solving compressible viscous flows, and provides a reference for further development of these methods toward engineering applications in fluid mechanics.

However, this study still has certain limitations. For example, only subsonic laminar flows are considered, while flows involving shock waves and turbulence are not solved. In the future, these challenges may be addressed by using advanced methods.

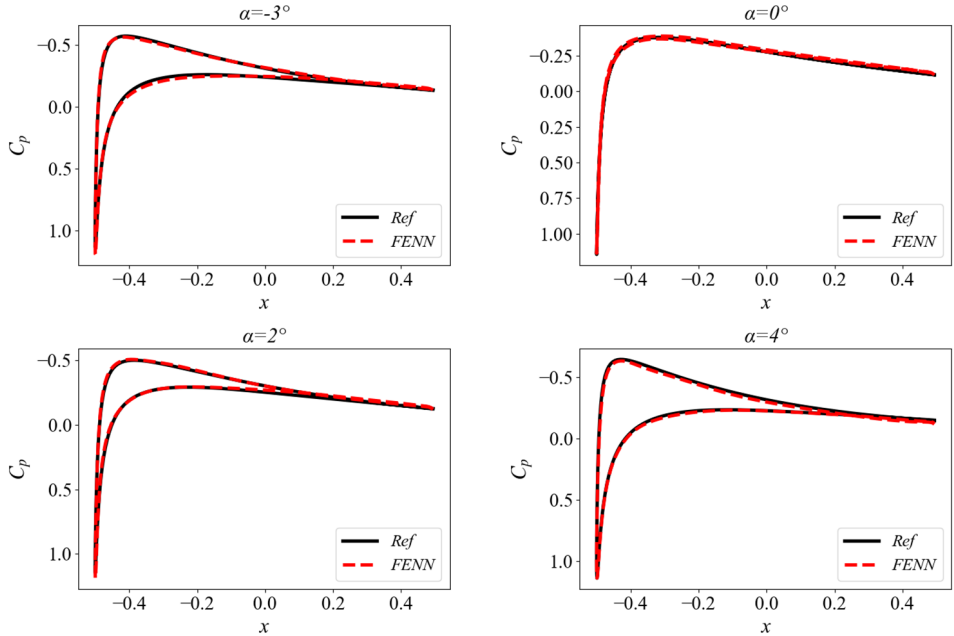

Figure 8. Pressure coefficients on the NACA0012 airfoil at different angles of attack obtained by the finite volume method (Ref) and FENN for solving the parametric problem involving compressible viscous flows around the NACA0012 airfoil.

## Acknowledgements

This research is supported by the National Natural Science Foundation of China (No. U2441211).

## Appendix A

Figure A1 shows the loss convergence of the feature networks for the shortest distance from the collocation points to the airfoil in numerical experiments as described in Section 3.

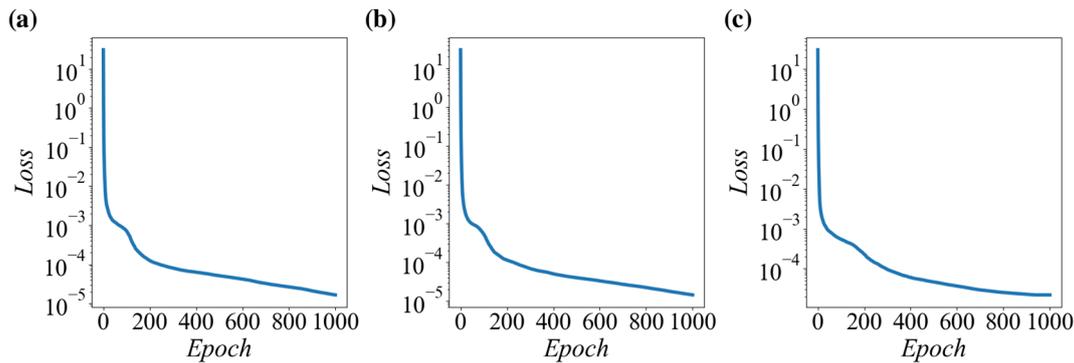

Figure A1. Loss functions of the feature networks. (a) NACA0012. (b) NACA2418. (c) NACA4424.